\documentclass[aps,prl,twocolumn,groupedaddress,floatfix]{revtex4}

\usepackage{graphicx}
\begin{document}

%\preprint{}

\title{Investigating Nearby Exoplanets via Interstellar Radar}


\author{Louis K. Scheffer}
%\email[]{SchefferL@janelia.hhmi.org}
%\thanks{}
\affiliation{Howard Hughes Medical Institute}


\date{\today}

\begin{abstract}
Interstellar radar is a potential intermediate step between passive observation of exoplanets and interstellar exploratory missions.
Compared to passive observation, it has the traditional advantages of radar astronomy.  
It can measure surface characteristics, determine spin rates and axes, provide
extremely accurate ranges, construct maps of planets,
distinguish liquid from solid surfaces, find rings and moons, and penetrate clouds.  
It can do this even for planets close to the parent star.
Compared to interstellar travel or probes, 
it also offers significant advantages.  
The technology required to build such a radar already exists, radar can return results 
within a human lifetime, and   
a single facility can investigate thousands of planetary systems.  
The cost, although high, is within the reach of Earth's economy, so it is cheaper as well.
\end{abstract}

%\keywords{}

\maketitle

\section{Introduction}
Radar astronomy has long been used to study the planets of the solar system\cite{ostro1993planetary}\cite{butrica1996see},
returning information difficult to obtain by optical methods.  
It was used to reconnoiter the planets, to find their ranges accurately, and otherwise
prepare for physical visits by probes.  
If the type of information returned by radar studies could be obtained for exoplanets, it would mark a huge increase in human knowledge.

However, at first glance, the idea of interstellar radar seems completely impractical.  With our biggest radio telescopes,
largest transmitters, and most sensitive receivers, we can still barely ping the moons of Saturn, about one light-hour away.  
Since all other stars are light {\it years} away, and the energy returned by an object at range R goes as $1/R^4$, 
clearly we are a long way from interstellar radar.  The gap in capability is about 20 orders of magnitude.

However, a look at the physics shows there are no fundamental barriers to achieving a gain of this magnitude, just issues of size and cost.  
Furthermore, a look at the technologies of phased array transmitters and receivers shows the costs are high but within the capabilities of Earth.  
This is something the human race could do now, if it was willing to expend military-budget class sums on the project.
In this sense it is much easier than interstellar travel.

In this article, we look at what it would take to build an interstellar radar, and what the possible advantages might be.
\section{Previous work}
Williams proposed an interstellar radar in 1985\cite{williams1985radar}, noting the only barriers were size and cost.
His suggested implementation of the transmitter was a fleet of 16,000 1-km sized spacecraft, 
each weighing 50,000,000 kg, all inside the orbit of Mercury.  
A similar array spread outside the orbit of Jupiter forms the receiever;
the huge baseline is used to resolve the planets interferometrically.  
Compared to the current work, this approach requires space-based construction well beyond the current state of the art, and would
doubtless be expensive in both construction and maintenance.  
Also, the calculations assumed a filled aperture transmitter when computing the beam width; 
this might be quite difficult with the antennas in orbit.

Rzhiga, also in 1985, similarly concluded the only barriers are size and cost, and calculated that a 70 km antenna with 
a 2 TW transmitter and 10 K receiver could detect reflections from Earth size planets around Alpha Centauri\cite{rzhiga1985radar}.  
No specific means of construction were proposed, but he suggested space based antennas to avoid problems with the troposphere. 
By contrast, in the current work the tropospheric
fluctuation are calibrated out, allowing operation from the Earth's surface.
Rzhiga also notes that the receiver array could allow interstellar transmission of television even with a modestly sized transmitting station
of 100 KW power and a 10 meter antenna.

\section{Implementation}
The current record holder for long distance radar observation is from Earth to Saturn's rings and its moon Titan.  
This is the upper limit of range for Arecibo, the largest facility,
since a more distant target will not stay in the beam long enough for a round trip signal\cite{campbell2003radar}\cite{muhleman1995radar}.

In this paper, we will work by scaling the properties of this facility (rather than using the raw radar equations) 
since this already takes a multitude of practical losses into consideration.
The basic parameters for Titan imaging are a one-way light travel time of 1 hour 7.5 minutes, 
a receiving antenna size of about 73,000 m$^2$,
and an equivalent isotropic radiated power (EIRP) of 20 TW ($2\cdot 10^{13}$\ watts), from
a transmitter power of 1 MW and antenna gain of 73 db\cite{renzetti1988}.

As a first estimate, assume we wish to increase the range of Arecibo by 100,000 times.  
This suffices to image a Titan sized moon or planet at 13 light years, or an Earth size planet 20 light years away, a distance that encompasses
at least 100 star systems\cite{recons2012}.
The required product of transmitter power increase and receiving area increase is then $10^{20}$.  
Using phased array technology, an $N$ element array has $N^2$ times the EIRP of a single element, but only $N$ times the collecting area, so
it's easier to build big transmitters than big receivers.  Roughly balancing costs implies
a transmitter $10^{15}$ times more powerful and a receiver $10^5$ times more sensitive, as shown below.

On the receiving side, the most straightforward approach is to build 100,000 Arecibos, 
for a total collecting area of about 7,300 square kilometers.  
Using modern electronics and time standards, the signals from all the antennas can be combined in phase.  
This is a routine operation in today's VLBI networks, 
and radar is in many ways a much easier case since only a narrow bandwidth will need to be correlated.  
The combined Arecibos would require an area of about 10,000 square kilometers, or an area 100 km on a side.  
This is certainly possible in the more remote desert areas - for example, a radio-quiet zone of the required size already exists in the interior of Australia\cite{acmaQuiet}.
However, it would be better to spread the antennas out, since this increases resolution.

Since Arecibo can only view a limited portion of the sky, 
a better approach might be an array of steerable dishes.  
The total area needed would be somewhat less than the multiple Arecibo case, 
since Arecibo does not use the whole aperture at any given time - 
some margin is needed so steering the feed does not cause the pattern to leave the dish.  
The active area is an oval 213 by 237 meters\cite{goldsmith1996second}, for an
area of about 40,000 m$^2$.  
The same collecting area would be obtained with an array of 6 million 30 meter antennas, or 36 million 12 meter dishes, or other combinations.
It is also easier to extend this approach to higher frequencies than Arecibo style technology.

On the transmit side, one possibility might be an array of $10^{15}$ small antennas,
each radiating a 10 mw signal, circularly polarized, into a half hemisphere.  
If correctly summed in phase, this generates an EIRP of $2\cdot10^{28}$ watts, about $10^{15}$ times the strength of Arecibo, as required.
Assuming a wavelength of 2 cm (15 GHz) and elements spaced $\lambda/2$ or 1 cm apart, the array is about 316 km on a side.
Such an array might be constructed from $10^{13}$ tiles, each about 10 cm on a side.  
The bottom of each tile has a metallic ground plane, and on the top are printed 100 antennas. 
In the center of the tile there is a single chip that transforms 2 watts of DC into 100 separate 10mw RF signals with 50\% efficiency.  
Each tile must know its location and be able to independently adjust the phase of each of its antennas, 
but these are tasks well within the capabilities of modern electronics\cite{scheffer2005scheme}.
The main technical challenge is maintaining phase coherency across the array, and a number of solutions have been proposed\cite{davarian2007uplink}.
Likely best for an array of this size would be a space based calibrator, operating within a few degrees of the target.   
It could both send to the array on nearby frequencies and monitor the transmitted phases, providing a common time reference and several
possible methods of calibration.

Powering such an array is an excellent application for solar power.  
It would be acceptable to run such a transmitter only during daylight hours and clear weather,
so there is no need for energy storage.  
The power can be utilized right where it is produced, so there is no need for distribution.
Finally, the electronics need low voltage DC, which is exactly what solar cells produce - there is no need for energy conversion.
Therefore for this analysis we'll assume the the 2 watts of DC comes from a 20\% efficient solar cell coating the tile.

The total amount of RF energy is about 10 TW, which (during the day) is roughly $1/10$ of the energy of the sunlight falling on the array.  
The outgoing beam is only 1/10 the intensity of sunlight and hence would not be instantly destructive 
to airplanes and satellites that might cross its path, though precautions would surely be advisable.
Unlike the receiver array, the transmitter should be as compact as possible; 
otherwise beam width, already rather small, becomes a serious limitation.  
Unlike the receiving array, there is no need to locate this facility in a radio-quiet zone.
\section{Cost}
What about the cost?  This section uses US\$ as of 2012, with suffixes M=million ($10^6$), B=billion ($10^9$), and T=trillion ($10^{12}$).

Arecibo cost \$9.7M to build in 1960, with upgrades of \$8M in 1974 and \$27M in 1997\cite{matthews2012}.
A similar but modernized telescope (FAST in China) is being constructed today, 
and is expected to cost about 600 million RMB, or about \$100M\cite{nan2008introduction}.
100,000 of these would cost about \$10T,
assuming enough suitable terrain can be found. (This style of telescope is most easily built in a pre-existing natural depression.)
Alternatively, 12 meter dishes are currently about \$300,000 each\cite{CSIROantennas}.  36 million of these would total about \$11T. 
A third alternative would be to replicate the SKA, 
which is aiming for a square kilometer of collecting area for a cost of about \$2B\cite{dewdney2009square}.
4000 of these would cost \$8T.

All of these prices are probably over-estimates, since no-one has yet tried ordering radio telescopes in bulk quantities suitable 
for mass production.
For manufactured items in general, each doubling of the volume typically reduces cost 
(for electronics and manufacturing) to somewhere between 90 and 95\% of the original cost\cite{yelle1979learning}\cite{malerba1992learning}.  
For the volumes considered here, with normal learning curves the cost per item will be several times cheaper by the end of the production run.

On the transmitter side, as of 2012, 
you can purchase toys for considerably less than \$10, more than 10cm
on a side, with significant RF electronics content.  
A tile is simpler than many of these toys - just a slab of plastic, metallized on one side, with the other containing a solar cell, a chip, and
an array of printed antennas.
With a bulk order of $10^{13}$ tiles, a cost of \$1 or less seems reasonable.
Installation could be as simple as strewing the tiles across a desert landscape - using GPS style triangulation, each tile could
determine its own location, orientation, and a common time reference.  It could then respond to broadcast commands telling what
signals to transmit.
Compared to the receiving side, the expected saving from mass production would be less on the transmitting side, 
since the most costly components, the solar cells and chips, are already in mass production.

Combining the transmitter and receiver gives a total cost of \$20T or  less.  
If built over a 10 to 20 year period, this is way above any current science budget, but is comparable to spending on
consumer electronics or militaries - the global electronics industry is about \$1T per year, 
and all defence spending combined is about \$1.2T per year.  
Alternatively, this would be a twenty year project at \$140/year for each person on earth, or \$1000/year for each person in a developed economy.
So humanity could do this project today if it felt the need.  
\section{Aiming and Listening}
There are two problems with interstellar radar that make accurate knowledge of positions and ranges mandatory.  
First, the beam has to hit the planet, and second we need to know when to expect the return.  

Hitting the target with the beam is non-trivial since the beam is small, a necessary consequence of the high flux density needed.
In the case proposed here, the transmitter is about $1.6 \cdot 10^7 \lambda$ on a side, so the natural beam width is about 0.013 arcseconds.
(Thus for the interstellar case, as well as within our solar system, radar is a poor search tool and much better suited to following up
existing discoveries.)
The location of the star on the sky, and its proper motion, must be known well enough to predict the location of the star when the
beam gets there.
Furthermore, we need to know where the planet is with respect to the star, since it can range up to hundreds of beam widths away.
A 1 AU orbit at 1 parsec range may appear up to 1 arcsecond away from the star, so the orbit must be known to roughly 1\% accuracy.  
In general, a combination of techniques will be needed to establish an adequate orbit - for example,
radial velocity, astrometric, or transit techniques alone will not give good enough positions.
The good news is that the beam is large enough to illuminate any moons or rings of the planet it is targeting.

In addition to the position, the distance to the system must be known quite accurately, for two reasons.  
The first requirement is to hit the planet, which is a moving target.
This requires knowledge of how long the signal will take to get there.  
The second reason, requiring yet more accuracy, is to know when to look for the return signal.  
This is potentially a more troubling problem, particularly if the receiving
array has rotated out of view when the signal returns.
There are at least 3 potential solutions to this problem.  
One possibility would be to build 3 facilities spaced around the world, as JPL does with the Deep Space Network, though this would be expensive.  
Alternatively, the receiver could be space based and capable of 24 hour operation, though this would be more expensive yet.
The simplest solution would be just to wait.
A typical radial velocity for a nearby star is about 10 km/second\cite{reid2007palomar}.  
At this rate, the return time will change at about 2000 seconds per year.  
Since the worst case is 24 hours off, or about 86400 seconds, we can get optimal reception within 40 years or so.  Unfortunately there will
be a few stars with both low radial velocities and inconvenient round-trip times for which waiting is not a practical solution.

In order to make sure the receiver is in view when the signal gets back, the distance to the star must be known to a few light hours.  
This will need to be established through parallax measurements, and at 13 light years, this implies a parallax accuracy of roughly 10 $\mu$as.  
This accuracy can be attained by Gaia technology\cite{klioner2010gaia}, 
though Gaia itself is not this accurate on stars this bright since they saturate the detectors\cite{gardiol2005gaia}.
Once an initial signal is received, uncertain ranges will no longer be a problem - 
the uncertainty in distance immediately drops from a few light hours to a light millisecond or less.  
That's an improvement of $10^8$ or so in one observation.
\section{Transmission}
Different radar techniques and modulations are strongly preferred depending on the rotational speed and size of a target\cite{harmon2002planetary}.
Sizes and rotation rates vary widely even within the solar system, from targets such as Mercury that are compact and 
rotate slowly, to asteroids which are small and rotate quickly, to Saturn's rings, which are large and orbit with high velocity.  
Within the solar system, different radar techniques are used for each target type.

For extrasolar planets, rotation rates, sizes, and the presence of rings and moons may all be unknown 
until we get the echos.  
Rather than waiting another round trip time
to try again, it probably makes sense to probe each planet with a set of ranging waveforms.  
These should be picked so at least one gives decent results over the range of expected planetary characteristics.

A single radar observation can reveal the target size from the range of delays.  It also shows the rotational velocity, projected onto the
line of sight, from the doppler shift, but it does not show the rotational axis, which is important for both theoretical studies of
solar system evolution and potential habitability.  However if features can be resolved this can be measured by repeated observation.
An accurate orbit of the planet around its parent star needs several separated
observations as well. 
Therefore rather than a single long observation, it may be better to observe a number
of times over a range of hours, days, weeks, months, or even years.  
This range of times (if the solar system is typical) should allow 
determination of accurate orbits and axial tilts.
\section{Strategy}
There are about 6-10 stars per light-year of distance within 15 light years of Earth.  
Each additional light year of distance requires 2 more years of round trip time.  
In perhaps the simplest scheme, the transmitter starts by illuminating the 
planets around the nearest star, then switches to the next further star when twice the distance between them (in light time) has elapsed.  
This keeps both the transmitter and receiver busy all the time.  For these nearby stars, several months per star will be available for analysis.

There is no reason, except economics, that this method could not be extended out further than the initial 13 light years.  
There is no high energy density to cause trouble with material properties,
and the only technologies needed are precision timing, accurate positioning, and mass production.  
A transmitter 3000km in size would have 100 times the number of transmitters, and 10,000 times the EIRP, leading to a range 10x greater.  
This would be about 40 parsecs or 130 light years, encompassing at least 6,000 known stars\cite{HYG}.

As the range increases, there will be more systems at similar distances, and hence less time per system.
A reasonable minimum might be one day per system, 
which will happen when the range is such that each additional 1/2 light year of radius incorporates 365 new systems.
This will happen somewhere about 100 light years, and past this range only a subset of targets can be examined.
\section{Advantages}
There are many advantages to radar observations.
When compared to interstellar probes or missions, one big advantage is that results will come within in a human lifetime.  For example, a radar
observation of $\tau$ Ceti will be complete 25 years after the signal is sent.  This is already less elapsed time than some deep space missions such as
Voyager, that are still returning data 35 years after their launch.  With a probe,
however, even at an optimistic 10\% of the speed of light, the probe will take 125 years to get there, and another 12 years to send the signal back.
Therefore any probe sent will only return data to an investigator's descendants.

Another advantage is that a radar facility can observe hundreds to thousands of nearby systems in detail, 
including many interesting targets such as $\tau$ Ceti (suspected to have 5 planets\cite{Tuomi2012}) and Epsilon Eridani.  Each probe, in contrast, will only observe one system.  To be fair however, some probe architectures, such as a large
launching facility that sends out lightweight ``starwisp'' style probes, offers similar benefits.

Radar observations can perform detailed studies of planets close to their suns, difficult by other means.  
In particular, optical methods have trouble analyzing close-in planets that do not eclipse their star, since
they are very difficult to resolve against the nearby and much brighter star.  This problem is sufficiently serious that even in
our own solar system, the spin properties of Mercury had to be determined by radar\cite{pettengill1965radar}.

Radar measurements will immediately improve ranges to other systems by many orders of magnitude.  
They will also improve velocity with respect to the solar system, though by not as much since a great deal of work has already gone into precise radial velocities.
Both the distance and the velocity will be very helpful when sending a probe.

Radar investigations of a planet will also show the presence of moons and/or rings.
If they are observed several different times the detailed orbits can be inferred.

Radar can also determine the rotation rates for all bodies, and their axis of rotation.
In the solar system, this was used to determine the rotation rates of Mercury and Venus - 
Mercury because it is close to the sun and hard to observe optically, and Venus because it is shrouded by clouds.  
Both of these situations are likely to occur in other solar systems.

Radar imaging can generate crude images of planets by means of ``delay-doppler'' mapping.  
The available resolution depends only on the signal to noise of the echo,
but not on the resolution of the beam or the distance to the planet.
Titan has been mapped with 10 degree resolution (hundreds of pixels)\cite{black2010global}, 
with the Moon\cite{zisk1974high}\cite{thompson1987high}, Venus\cite{rogers1969venus}, Mercury\cite{butler1993mercury}, and Mars\cite{harmon1999mars} mapped to much higher precision.  
Obtaining even low resolution maps for exoplanets will be difficult with other techniques.  
The maps obtained, however, will still be North-South ambiguous, for at this range interferometric disambiguation\cite{butrica1996see} will not be possible.

Another advantage of radar is that it illuminates the planet with EM waves of known polarization,
and hence reflections of the same and opposite polarizations can be distinguished. 
These reflections, and their differences, allow various physical properties to be inferred,
such as surface roughness on the scale of a wavelength or greater,
or potential dielectric constants of the reflecting material\cite{simpson1992radar}.
This type of analysis has been used (along with orientation information) 
to to infer the presence of ice at the poles of the Moon\cite{stacy1997arecibo} and Mercury\cite{harmon2001high}, and investigate
properties of the ice-caps of Mars\cite{harmon1999mars}.   
Radar reflections can also distinguish liquid from solid surfaces\cite{campbell2003radar}.

For SETI, a potential advantage is the prospect of detecting civilizations that are not trying to communicate, 
and perhaps ones that have self destructed, depending on the durability of their structures.
Cities, for example, are well known to have much brighter radar echos than natural surfaces, 
since they contain many flat surfaces at right angles, a 
geometry that contributes to strong radar reflections\cite{long1975radar}.  
They also exhibit several other ``unnatural'' reflectivity properties, since
as strong reflections at incidences grazing the surface (due to vertical walls)
 and stronger reflections when aligned with the compass directions\cite{kayton1997avionics} (for Earth cities, anyway).

Another possibility is that civilizations could leave very long lasting
and easy to find radar artifacts.  Radar reflectors are one of the few 
examples where objects of size typical of human construction may be visible
across interstellar distances.  The radar cross section of a trihedral (corner reflector) scales as the fourth power of the side length $a$:
$$\sigma = \frac{12\pi a^4}{{\lambda}^2}$$
A trihedral reflector 120 meters on a side would provide a 15 GHz radar reflection as strong as that of a 5000 km diameter moon.  
If this trihedral was cut into a planet's moon, this echo would blink on and off as the moon rotates, 
an effect that would surely draw attention to its likely artificial nature.
An advantage of such an artifact is that 
no aiming or maintenance are required, and it could outlast the civilization that created it by many millions of years.
\section{Other uses}
There are many other possible uses for a system of this size.  Clearly the receiver portion would make a great radio telescope.  The transmitter
part can be used to blast debris out of Earth orbit, or provide power to far away probes.

The radar could image any object within the solar system, Kuiper belt, or Oort cloud, 
though an alternative receiver would need to be used for some particular light travel times,
for which the echo returns when the main receiver is out of view.

The transmitter would make a fine interstellar beacon, if Earth chooses to use it that way.  
It outshines the sun by a factor of 25 - the Sun has an EIRP of $4\cdot10^{26}$ watts.  
The flux per bandwidth of a CW radar transmission as described here, at 10 light years range, is roughly 20,000,000,000,000,000,000 Jansky.  
A radio telescope would not need to be pointed at the transmitter to see this signal;
it would not even need to be pointed anywhere in the general direction.  
Unless it is unusually well shielded, simply being turned on and on the right side of the planet should suffice.
On the other hand, for any particular target of the radar investigation, the signal will be highly intermittent, and may never re-occur.

If after a radar investigation, Earth sends a probe or probes, the facility can serve as uplink and downlink.
\section{Conclusions}
Interstellar radar can give considerable information about nearby exoplanets that cannot be obtained through optical means.  It is within the range of Earth's technology and economy even now.  Though quite expensive, it is surely cheaper than an interstellar probe, and would make an excellent intermediate step.
\bibliography{radar}

\end{document}